\documentclass[10pt,twocolumn,letterpaper]{article}

\usepackage{cvpr}              % To produce the CAMERA-READY version
\usepackage{graphicx}
\usepackage{amsmath}
\usepackage{amssymb}
\usepackage{booktabs}
\usepackage{pifont}
\usepackage{subcaption} % 或者 \usepackage{subfig}
\usepackage[pagebackref,breaklinks,colorlinks]{hyperref}
\usepackage{xcolor}
\usepackage{multirow}

\usepackage[capitalize]{cleveref}
\crefname{section}{Sec.}{Secs.}
\Crefname{section}{Section}{Sections}
\Crefname{table}{Table}{Tables}
\crefname{table}{Tab.}{Tabs.}
\crefname{table}{Tab.}{Tabs.}

\def\paperID{7340} % *** Enter the Paper ID here
\def\confName{CVPR}
\def\confYear{2026}

\usepackage[x11names]{xcolor} 
\usepackage{colortbl}  
\definecolor{tablecolor}{RGB}{230,230,230}
\definecolor{impgre}{RGB}{33,166,102}
\definecolor{imp}{RGB}{192,0,0}

\begin{document}

\title{
AINet: Anchor Instances Learning for Regional Heterogeneity in Whole Slide Image
}

\author{Tingting~Zheng\textsuperscript{\rm 1}\quad Hongxun~Yao\textsuperscript{\rm 1*}\quad Kui~Jiang\textsuperscript{\rm 1} \quad Sicheng~Zhao\textsuperscript{\rm 2}\quad Yi~Xiao\textsuperscript{\rm 3}\\
\vspace{-1mm}
 {\textsuperscript{\rm 1} Harbin Institute of Technology}, {\textsuperscript{\rm 2}Tsinghua University}, {\textsuperscript{\rm 3}Zhengzhou University}\\
}

\maketitle

\begin{abstract}
Recent advances in multi-instance learning (MIL) have witnessed impressive performance in whole slide image (WSI) analysis. However, the inherent sparsity of tumors and their morphological diversity lead to obvious heterogeneity across regions, posing significant challenges in aggregating high-quality and discriminative representations. To address this, we introduce a novel concept of anchor instance (AI), a compact subset of instances that are representative within their regions (local) and discriminative at the bag (global) level. These AIs act as semantic references to guide interactions across regions, correcting non-discriminative patterns while preserving regional diversity. Specifically, we propose a dual-level anchor mining (DAM) module to \textbf{select} AIs from massive instances, where the most informative AI in each region is extracted by assessing its similarity to both local and global embeddings. Furthermore, to ensure completeness and diversity, we devise an anchor-guided region correction (ARC) module that explores the complementary information from all regions to \textbf{correct} each regional representation. Building upon DAM and ARC, we develop a concise yet effective framework, AINet, which employs a simple predictor and surpasses state-of-the-art methods with substantially fewer FLOPs and parameters. Moreover, both DAM and ARC are modular and can be seamlessly integrated into existing MIL frameworks, consistently improving their performance.
\end{abstract}

\section{Introduction }
\label{Instroduction}

\begin{figure}[!t] \centering
  \includegraphics[width=0.48\textwidth]{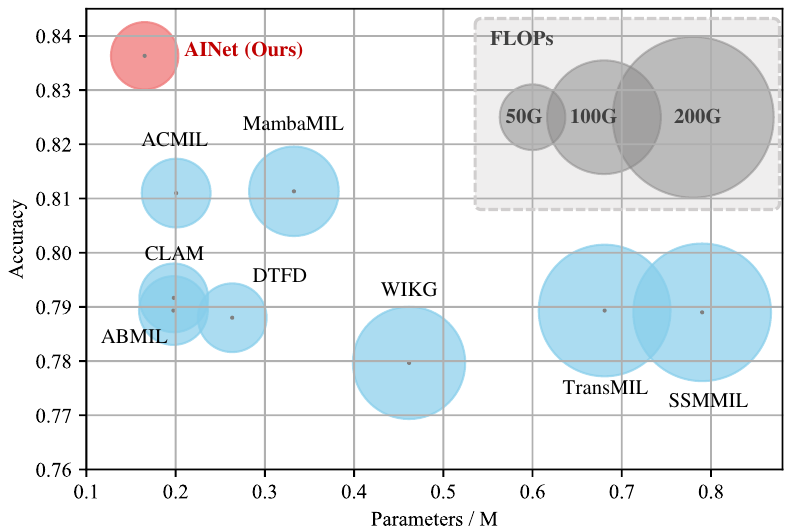}
\caption{Comparison results of  FLOPs (G), parameters (M), and average accuracy with representative MIL methods on TCGA-ESCA, TCGA-BRCA, and BRACS datasets. AINet delivers superior performance with improved efficiency.} 
\vspace{-5mm} 
\label{fig_FLOPs}
\end{figure}

Histopathology slides provide cellular morphology and tissue microenvironment, which are indispensable for accurate cancer diagnosis~\cite{sun2025cpathcvpr,quan2024globalPR}.  
With the rapid advancement of digital pathology, physical tissue slides can now be digitized into high-resolution whole slide images (WSIs), opening new avenues for applying deep learning to tumor diagnosis~\cite{hou2016patch,raswa2025histofscvpr,redundancy_cvpr2024MIL}. Nevertheless, the inherent gigapixel scale of WSIs, coupled with the sparse and scattered distribution of tumor regions, makes pixel-level manual annotation highly expensive and time-consuming~\cite {li2024generalizablecvpr,2020MSDAMIL_cvpr}. To alleviate these limitations, multi-instance learning (MIL) has emerged as a promising paradigm, treating a WSI as a ``bag" of instances (patches) and leveraging bag-level labels to guide feature learning~\cite{RankMix,shi2024vilacvpr}.

\begin{figure*}[t] \centering
  \includegraphics[width=0.96\textwidth]{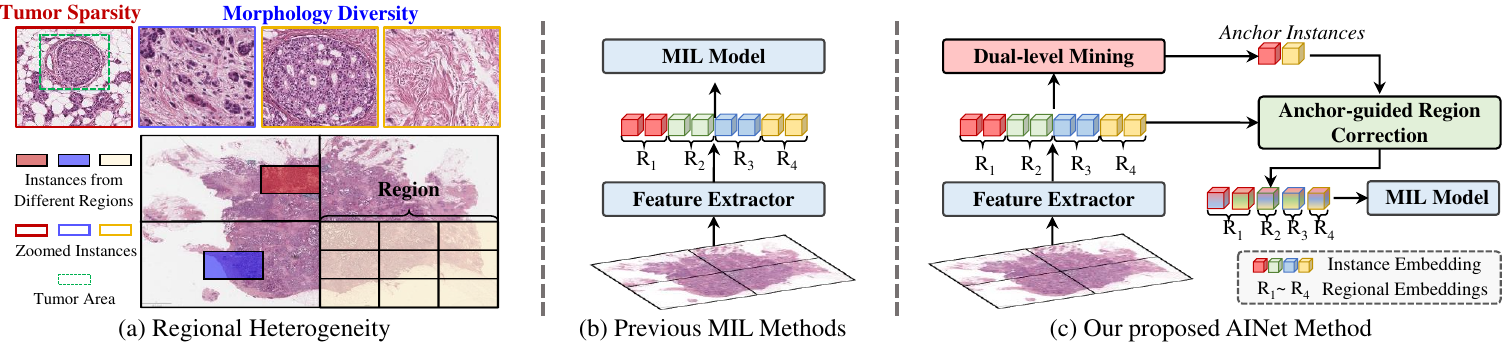}
\caption{(a) Regional heterogeneity within a WSI originates from tumor sparsity and morphological diversity. (b) Previous MIL methods often neglect this heterogeneity, resulting in suboptimal feature aggregation. (c) The proposed AINet alleviates this limitation via DAM and ARC modules, yielding more discriminative representations and improved predictive accuracy. } 
\vspace{-5mm} 
\label{fig_mot1}
\end{figure*}

Despite recent progress, MIL-based approaches for WSI analysis continue to be fundamentally limited by intra-WSI heterogeneity, a key challenge illustrated in Figure~\ref{fig_mot1}(a). 
This manifests in two primary forms: \textbf{(i)} the sparse distribution of tumor regions, which obscures discriminative instances within a vast number of non-informative ones, 
and \textbf{(ii)} the significant morphological diversity among instances, which hinders the aggregation of a comprehensive and coherent global context. 
Currently, existing methods designed to address these challenges can be broadly categorized into three paradigms: instance-, bag-, and pseudo-bag-level approaches~\cite{CLAM,2022DTFDMIL_CVPR}.

Instance-level methods predict labels for individual instances before aggregating them for a bag-level prediction~\cite{2019RNNMIL_Nat}.
However, in WSIs where tumor regions are sparse and non-lesion instances dominate, these methods often focus on uninformative instances and overlook global context, resulting in suboptimal representations~\cite{2024CVPRWIKG,2023ILRA_MIL}. Bag-level methods aim to capture global context through advanced architectures, including attention mechanisms~\cite{ABMIL}, graph neural networks~\cite{2024CVPRWIKG,HGNN_graph}, and Transformers~\cite{Transmil,tang2024featureRRT}, which aggregate multiple instances into a unified bag-level representation.
Despite these sophisticated designs, these approaches are computationally expensive and often fail to produce high-quality representations on highly heterogeneous instance sets, as aggregation can be diluted by uninformative or conflicting features.

Decomposing WSIs into multiple regions (or pseudo-bags) has emerged as an effective strategy for handling instance heterogeneity in MIL~\cite{2022DTFDMIL_CVPR}. However, due to sparse tumors and uneven feature distributions, many regions lack informative lesions or diversity cues, leading to representations that are neither sufficiently discriminative nor comprehensive~\cite{ProtoDiv,liu2023pseudoMixup}. To address this issue, cross-region fusion has been widely adopted to enhance feature representations by aggregating complementary information. Yet, the heterogeneity across regions may also introduce redundant or noisy features, thereby undermining its effectiveness~\cite{cvpr24PAMIL}. Consequently, a critical challenge remains: \emph{to devise a concise yet effective solution that mitigates regional heterogeneity by enhancing the discriminativeness of all regions while conserving their diversity and representativeness.}

Drawing inspiration from prototype learning, complex distributions can be effectively characterized by a compact set of reference points (called anchors)~\cite{ge2024beyondanchor}. 
Beyond representation, anchors act as semantic bridges~\cite{ge2024beyondanchor}, facilitating the correction of heterogeneous data and the discovery of discriminative features \cite{rothenhausler2021anchor, tan2022fedproto, qin2024discriminative}. Motivated by these insightful properties, we introduce a novel concept, i.e., anchor instance (AI), to elegantly address heterogeneity across regions. 
Specifically, AIs constitute a compact set of representative instances selected from all regions, which serve as semantic reference points to guide regional interactions, thereby correcting the uninformative regions induced by tumor sparsity and morphological diversity. The AI paradigm directly addresses the aforementioned requirement by: \textbf{1) providing stronger, globally salient features} that enhance discriminative power for region correction, and \textbf{2) summarizing representative information from each region} to guide the interaction of diverse patterns, thereby preserving diversity while mitigating heterogeneity.

\begin{figure*}[!t]
\centering
\includegraphics[width=0.96\textwidth]{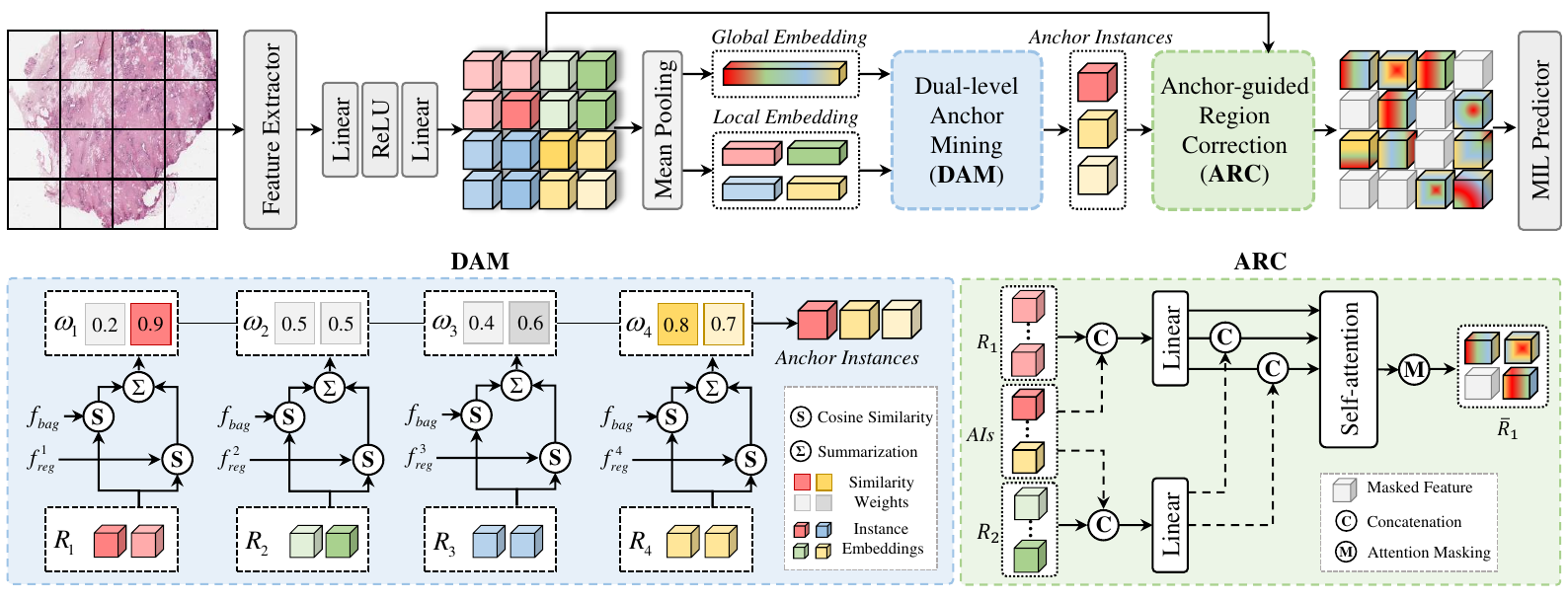}
\caption{Overview of the proposed AINet for WSI classification. Each bag is first divided into multiple spatial regions. The dual-level anchor mining (DAM) module identifies a compact set of anchor instances (AIs), which are then incorporated by the anchor-guided region correction (ARC) module into regional representations to guide the aggregation of heterogeneous features.}
\vspace{-3mm}
\label{AINet}
\end{figure*}

To identify AIs, we first design a dual-level anchor mining (DAM) module. Guided by the requirement that AIs should be globally discriminative and locally representative, DAM quantifies the salience of each instance by computing the similarity to both a global bag embedding and its local regional embedding. The dual similarity scores are combined to rank all instances, and the top-ranked ones are selected as the final AIs. To effectively utilize the AIs, we further devise an anchor-guided region correction (ARC) module. Rather than a simple assignment or fusion, ARC employs AIs as semantic pivots to facilitate interaction between heterogeneous regions. This design ensures that the corrected features maintain strong discriminative power and comprehensive representativeness, addressing the heterogeneity issue. Integrating the DAM and ARC modules forms our complete AINet framework. Using only a simple predictor, AINet outperforms state-of-the-art (SOTA) MIL-based models (Figure~\ref{fig_FLOPs}), underscoring the efficacy and efficiency of the anchor instance learning paradigm.

Our main contributions are summarized as follows: 
\begin{itemize} 
\item We introduce a novel concept of AI and develop AINet, a concise yet effective framework that extracts representative and discriminative instances for better WSI classification.
\item We design a DAM module to select AIs by comprehensively leveraging local (region) and global (bag) relationships, and an ARC module to mitigate heterogeneous interference during regional correction.
\item Extensive experiments on public datasets show that AINet outperforms state-of-the-art MambaMIL and PAMoE methods, achieving average accuracy gains of 2.50\% and 4.65\% on TCGA-BRCA, TCGA-ESCA, and BRACS, respectively, with lower FLOPs and parameters.
\end{itemize}

\section{Related Work}
\label{sec:Rew}
\subsection{Multi-instance Learning}
\label{sec:Rew_MILWSI}
MIL is a weakly supervised paradigm that models each sample as a bag of instances, where only the bag-level label is provided~\cite{ABMIL}. The assumption is that a bag is positive if at least one instance is positive, and negative otherwise~\cite{2019RNNMIL_Nat}.  MIL has been widely applied across various domains, including computer vision~\cite{shao2023videoMIL} and medical data analysis~\cite{guo2025focuscvpr,li2025advancingcvpr,wu2025learningcvpr}. To facilitate computer-aided diagnosis, MIL has gained significant attention in computational pathology for WSI analysis, driving advances in cancer subtype classification~\cite{cvpr24PAMIL}, survival prediction~\cite{shou2025graphiccv,ganguly2025mergegraphcvpr}, and lesion localization~\cite{liu2018landmarkMIL}. However, due to the extreme sparsity of tumor instances in gigapixel WSIs, learning discriminative bag-level representations remains highly challenging~\cite{ABMIL}. To overcome these limitations, some researchers have explored graph networks~\cite{behrouz2024graphmamb_ACM,zheng2025graphmamba}, attention mechanisms~\cite{zheng2023learning,zheng2025oodml}, Transformers~\cite{cvpr24PAMIL}, and Mamba~\cite{zheng2025m3amba,zhang20252dmambacvpr,zheng2025gmmamba} to predict or aggregate instance-level information for accurate bag-level prediction. While effective, their performance is often compromised when dealing with complex and heterogeneous tissue structures. To address these issues, we propose AINet, which captures a compact set of informative instances to guide the fusion of heterogeneous features, enabling high-quality and discriminative bag-level representations.

\subsection{Instance Selection in WSI classification}
Instance selection plays a crucial role in MIL-based WSI analysis, aiming to identify the most salient instances for accurate bag-level prediction~\cite{ABMIL}. Early studies mainly relied on global attention mechanisms, which assign attention weights to each instance~\cite{2019RNNMIL_Nat,CLAM}. However, a single attention operator can be inaccurate~\cite{shapleyval}.
To alleviate this issue, recent studies have introduced multi-branch attention mechanisms to capture instances with distinct morphological patterns, while others employ multi-head attention to model complex instance relationships for more reliable instance selection~\cite{Transmil,zhang2023ACMIL}. Despite these advances, relying solely on bag-level supervision and attention scores limits the diversity of selected instances~\cite{MHIM}. Several works have further explored similarity-based clustering, pseudo-labeling methods, and knowledge distillation to guide instance selection~\cite{BackWSIfinetuing23,shapleyval,AAAI2024CIMIL_pseudolabel_thre}. However, these methods often fail to select instances that are globally discriminative or locally representative, as they ignore the heterogeneity across WSI regions.
To this end, our proposed DAM effectively captures instances that are representative within diverse regions while being discriminative toward bag-level semantics, facilitating instance aggregation.

\section{Method}
\label{Method}
This section first formulates the MIL problem, then outlines the AINet framework, followed by detailed descriptions of the DAM and ARC modules.

\subsection{MIL Formulation}
Given a WSI $X$, the non-background tissue regions are divided into $N$ instances $\{(x_i, y_i) \mid 1 \leq i \leq N\}$, forming a bag. Here, $Y$ denotes the label of $X$, while the instance-level labels $y_i$ are unknown. 
The MIL formulation is expressed as:
\begin{align}
    \label{MIL_Formulation}
    Y_{i} &= 
    \begin{cases} 
        0, & \text{iff } \sum y_{i} = 0, \quad y_{i} \in \{0,1\}, \\ 
        1, & \text{otherwise}, 
    \end{cases}
\end{align}
In this case, a bag is labeled negative ($0$) if all instances are negative, and positive ($1$) otherwise.

Typically, the MIL framework for WSI analysis embeds each instance $x_i$ into a 1D feature vector $f_\text{ins}^i$~\cite{CLAM, 2022DTFDMIL_CVPR, cvpr24PAMIL}. After that, a scoring function then evaluates each feature to select the most informative instances or perform score-weighted aggregation for bag-level prediction. However, due to the heterogeneous tissue patterns in WSIs, such naive selection or aggregation often struggles to capture representative and discriminative representations. 

To alleviate this issue, we introduce a novel concept the AI. We identify a small subset of instances that serve as reference points to guide interactions across heterogeneous regions, thereby enabling the learning of high-quality and discriminative representations within a concise and effective framework.

\subsection{Overview of AINet}
Figure~\ref{AINet} presents an overview of the proposed AINet framework, which aims to capture a compact set of anchor instances and guides heterogeneous region correction to achieve high-quality and discriminative representations.

First, we denote a bag as a set of $ N$ instance features $\{f_\text{ins}^i\}_{i=1}^N$. We then divide the bag into $L = \frac{N}{Z}$ regions according to the spatial distribution (coordinates) of instances in the WSI, where $Z$ is the number of instances per region.

Next, the DAM module is employed to estimate instance importance by jointly comparing each instance with the region-level embeddings and the global-level embedding of the entire bag, making the selection of AIs $\{f_\text{AI}^t\}_{t=1}^T$ both locally representative and globally discriminative, where $T$ denotes the number of AIs and $t$ indexes the $t$-th AI. 
To address regional heterogeneity and correct the non-discriminative regions, ARC employs AIs $\{f_\text{AI}^t\}_{t=1}^T$ to guide cross-region interactions, enabling discriminative cues from AIs to refine the uninformative regions while preserving the diversity and representativeness of regional instances.
It is worth noting that a mask-based attention mechanism is incorporated into ARC to filter out redundant computations induced by massive instance interactions, thus resulting in compact bag representations.

Finally, the remaining instances are fed into an attention-based predictor to produce the final bag prediction $\hat{Y}$.

\subsection{Dual-level Anchor Mining}
To effectively identify sparsely distributed AIs while maintaining computational efficiency, each instance feature is first projected into a latent feature space through a lightweight multilayer perceptron (MLP), which consists of two linear layers and a ReLU activation. This projection produces $L$ regions of instance features $R_l = \{f_l^z\}_{z=1}^{Z}$, where $f_l^z$ denotes the $z$-th instance feature in the $l$-th region. We then compute two hierarchical embeddings to provide dual-level semantic guidance: a bag-level embedding $f_\text{bag}$ obtained by mean pooling (\text{MP}) over all instances within the WSI (i.e., the entire bag), and a region-level embedding $f_\text{reg}^l$ obtained by mean pooling over instances within each region $R_l$. This process can be formulated by:
\begin{equation}
    f_{\text{bag}} = \text{MP}(\{R_l\}_{l=1}^L), \{f_{\text{reg}}^l\}_{l=1}^L = \{\text{MP}(R_l)\}_{l=1}^L.
\end{equation}

As illustrated in Figure~\ref{AINet}, the DAM module computes the cosine similarity weights $w_l^z$ between each $f_l^z$ and both the region-level embedding $f_\text{reg}^l$ and the bag-level embedding $f_\text{bag}$, which means:
\begin{equation}
    w_l^z = 
\alpha \cdot \cos(f_l^z, f_{\text{reg}}^l)
+ (1-\alpha) \cdot \cos(f_l^z, f_{\text{bag}}), \label{eq:similarity_weight}
\end{equation}
where $\cos(\cdot)$ denotes the cosine similarity function, and $\alpha$ is set to 0.7 to balance local and global similarities. The similarity weights are then ranked, and the top $k\%$ instances with the highest scores are selected as anchor instances $\{f_{\text{AI}}^t\}_{t=1}^T$:
\begin{equation}
\{f_{\text{AI}}^t\}_{t=1}^T =
\operatorname{Top}_k (\{f_l^z, w_l^z\}_{z=1}^Z \,\big|\, l=1,\dots,L ),
\label{eq:ai_selection_full}
\end{equation}
where $\operatorname{Top}_k(\cdot)$ selects the top $T = k\% \times N$ instance features and $N = L \times Z$.

\subsection{Anchor-guided Region Correction}
With the DAM module, we obtained AIs that are both locally representative and globally discriminative.  
The proposed ARC module then aims to leverage these AIs as guidance to mitigate regional heterogeneity during cross-region interaction and correction. This approach allows us to use the AIs as explicit references, resulting in discriminative representations for more accurate predictions. Specifically, ARC incorporates the AIs into each region to form $f_\text{AIR}^l$, which is then linearly projected using $W_q^l, W_k^l, W_v^l$ into the query $Q_l$, key $K_l$, and value $V_l$.

Different from vanilla self-attention mechanisms, each $Q_l$ attends to Key and Value not only from the current $l$-th region but also from the adjacent $(l+1)$-th region, introducing more diverse and discriminative regional representations for interaction. The anchor-guided region correction process can be summarized as follows:
\begin{align}
&f_\text{AIR}^l  =[\{f_{\text{AI}}^t\}_{t=1}^T;\{f_l^z\}_{z=1}^Z],\\
&Q_l=W_q^lf_\text{AIR}^l,\,K_l=W_k^lf_\text{AIR}^l,\, V_l=W_v^lf_\text{AIR}^l,\\
&R_\text{cross}^l = \text{Softmax}(\frac{Q_l [K_l;K_{l+1}]^T}{\sqrt{d_K}})[V_l;V_{l+1}],
\label{eq7}
\end{align}
where $[\,\cdot\, ,\, \cdot\,]$ denotes the concatenation operation. 

While $R_\text{cross}^l = \{f_\text{cross}^{l,j}\}_{j=1}^{T+Z}$ encodes discriminative cues across regions and AIs, correcting a large number of instances inevitably introduces redundant information. To further enhance the quality of the corrected features, we apply an attention-based masking mechanism, which masks $M = r \times (T+Z)$ instance features with the lowest attention scores within each region, denoted as $\text{Mask}_r(\cdot,\cdot)$. Here, $r$ is the mask ratio.  
This procedure is described as follows:
\begin{align}
 & \{A^{l,j}\}_{j=1}^{T+Z} = \text{Attention}\left(\{f_\text{cross}^{l,j}\}_{j=1}^{T+Z} \right),\\
 &\{\bar{f}_\text{cross}^{l,j}\}_{j=1}^{T+Z-M} = \text{Mask}_r\left(\{f_\text{cross}^{l,j},A^{l,j}\}_{j=1}^{T+Z}\right).
\end{align}
Finally, the compact set $\bar{R}_l = \{\bar{f}_\text{cross}^{l,j}\}_{j=1}^{T+Z-M}$ is passed to a MIL predictor to achieve the prediction. In this work, we adopt a standard attention-based predictor, generating both region- and bag-level prediction results, i.e., $\hat{y}^l$ and $\hat{Y}$.

\subsection{Model Optimization}
To preserve essential information during feature projection before the DAM module, we apply a mean-squared error (MSE) consistency loss between the original instance features $\{f_\text{ins}^{l,z}\}_{z=1}^{Z}$ and their projected counterparts $\{f_l^z\}_{z=1}^{Z}$.  
In addition, cross-entropy losses are computed at both the region and bag levels to ensure stable optimization. The overall loss function is formulated as:
\begin{align}
&\mathcal{L}_\text{MSE} = 
\frac{1}{L\cdot Z} \sum_{l=1}^{L} \sum_{z=1}^{Z} 
\left\| f_\text{ins}^{l,z} - f_l^z \right\|_2^2,\\
 & \mathcal L_\text{region} \!=\! - \frac{1}{L} \sum_{l=1}^{L} \left[ Y \log \hat{y}^l + (1 - Y) \log(1 - \hat{y}^l) \right],\\
 & \mathcal{L}_\text{bag} = - \left[ Y \log \hat{Y} + (1 - Y) \log (1 - \hat{Y}) \right].
\end{align}
Finally, the overall loss function is defined by the following combination:
\begin{align}
 \mathcal L_\text{AINet} = \mathcal{L}_\text{bag}+ \mathcal L_\text{region}  +\mathcal{L}_\text{MSE}. 
\end{align}

\begin{table*}[!t]
\centering
\setlength{\tabcolsep}{5pt} 
\renewcommand{\arraystretch}{1} 
\newcommand{\Ft}[1]{\textcolor{red}{\textbf{#1}}}
\newcommand{\Sd}[1]{\textcolor{blue}{\textbf{#1}}}
\caption{Quantitative results of AINet compared with representative MIL baselines on TCGA-BRCA and TCGA-ESCA datasets.}
\begin{tabular}{llcccccc}
\toprule
\multicolumn{2}{c}{\multirow{2}{*}{Methods}} & \multicolumn{3}{c}{TCGA-BRCA} & \multicolumn{3}{c}{TCGA-ESCA} \\
\cmidrule(lr){3-5}  \cmidrule(lr){6-8}
\multicolumn{2}{c}{} & Accuracy & AUC & F1 & Accuracy & AUC & F1 \\
\midrule
\multirow{14}{*}{\rotatebox{90}{ResNet18}}
&ABMIL~\cite{ABMIL}	&0.862$\pm$0.025	&0.882$\pm$0.033	&0.915$\pm$0.015	&0.815$\pm$0.129	&0.882$\pm$0.126	&0.852$\pm$0.101	\\
&DSMIL~\cite{CVPR21DSMIL}	&0.823$\pm$0.021	&0.820$\pm$0.033	&0.892$\pm$0.014	&0.712$\pm$0.085	&0.823$\pm$0.078	&0.773$\pm$0.074	\\
&CLAM~\cite{CLAM}	&0.865$\pm$0.020	&0.890$\pm$0.029	&0.917$\pm$0.014	&0.821$\pm$0.091	&0.916$\pm$0.052	&0.844$\pm$0.085	\\
&TransMIL~\cite{Transmil}	&0.847$\pm$0.021	&0.846$\pm$0.036	&0.905$\pm$0.013	&0.815$\pm$0.081	&0.878$\pm$0.086	&0.847$\pm$0.065	\\
&DTFD-MaxM~\cite{2022DTFDMIL_CVPR}	&0.816$\pm$0.023	&0.810$\pm$0.033	&0.885$\pm$0.013	&0.821$\pm$0.067	&0.914$\pm$0.075	&0.851$\pm$0.057	\\
&DTFD-AFS~\cite{2022DTFDMIL_CVPR}	&0.823$\pm$0.028	&0.824$\pm$0.034	&0.892$\pm$0.017	&\Sd{0.865$\pm$0.074}	&0.912$\pm$0.063	&0.881$\pm$0.069	\\
&MHIM-ABMIL~\cite{MHIM}	&0.858$\pm$0.004	&0.883$\pm$0.020	&0.912$\pm$0.001	&0.828$\pm$0.119	&0.911$\pm$0.098	&0.866$\pm$0.085	\\
&MHIM-Trans~\cite{MHIM}	&0.848$\pm$0.022	&0.872$\pm$0.013	&0.905$\pm$0.012	&0.853$\pm$0.061	&0.911$\pm$0.044	&0.879$\pm$0.049	\\
&ILRA~\cite{2023ILRA_MIL}  	&0.857$\pm$0.035	&0.886$\pm$0.026	&0.908$\pm$0.027 	&0.803$\pm$0.136	&0.896$\pm$0.097	&0.838$\pm$0.116	\\
&SSMMIL~\cite{23MICCAIS4MIL_mamba}	&0.863$\pm$0.006	&0.896$\pm$0.032	&0.916$\pm$0.005	&0.783$\pm$0.111	&0.892$\pm$0.081	&0.823$\pm$0.096	\\
&ACMIL~\cite{zhang2023ACMIL} &\Sd{0.869$\pm$0.017}	&\Sd{0.900$\pm$0.019}	&\Sd{0.920$\pm$0.009}	&0.866$\pm$0.051	&0.932$\pm$0.044	&\Sd{0.893$\pm$0.037}	\\
&WIKG~\cite{2024CVPRWIKG} 	&0.863$\pm$0.018	&0.887$\pm$0.030	&0.914$\pm$0.013 	&0.770$\pm$0.131	&0.874$\pm$0.106	&0.793$\pm$0.121	\\
&MambaMIL~\cite{Hao2024mambamil}	&0.868$\pm$0.017	&0.878$\pm$0.032	&0.917$\pm$0.009	&0.860$\pm$0.063	&\Sd{0.944$\pm$0.050}	&0.886$\pm$0.044	\\
&PAMoE~\cite{wu2025learningcvpr} &0.853$\pm$0.017	&0.854$\pm$0.034	&0.908$\pm$0.011 &0.800$\pm$0.098	&0.897$\pm$0.090	&0.836$\pm$0.086\\
\rowcolor{tablecolor} 
&\textbf{AINet (Ours)}	
&\Ft{0.878$\pm$0.016}	&\Ft{0.901$\pm$0.028}	&\Ft{0.924$\pm$0.011}	&\Ft{0.873$\pm$0.098}	&\Ft{0.946$\pm$0.046}	&\Ft{0.894$\pm$0.080}\\
\midrule
\multirow{14}{*}{\rotatebox{90}{PLIP}}
&ABMIL~\cite{ABMIL}	&0.870$\pm$0.019	&0.900$\pm$0.010	&0.918$\pm$0.012	&0.923$\pm$0.042	&0.976$\pm$0.029	&0.934$\pm$0.036	\\
&DSMIL~\cite{CVPR21DSMIL}	&0.856$\pm$0.019	&0.895$\pm$0.013	&0.911$\pm$0.012	&0.904$\pm$0.067	&0.961$\pm$0.037	&0.919$\pm$0.057	\\
&CLAM~\cite{CLAM}	&\Sd{0.874$\pm$0.021}	&0.895$\pm$0.013	&\Sd{0.922$\pm$0.012}	&0.917$\pm$0.060	&0.973$\pm$0.018	&0.929$\pm$0.050	\\
&TransMIL~\cite{Transmil}	&0.859$\pm$0.026	&0.888$\pm$0.019	&0.912$\pm$0.017	&0.904$\pm$0.062	&0.978$\pm$0.018	&0.920$\pm$0.051	\\
&DTFD-MaxM~\cite{2022DTFDMIL_CVPR} 	&0.862$\pm$0.019	&0.861$\pm$0.014	&0.916$\pm$0.010	&0.911$\pm$0.051	&\Sd{0.981$\pm$0.020}	&0.927$\pm$0.039	\\
&DTFD-AFS~\cite{2022DTFDMIL_CVPR} 	&0.840$\pm$0.012	&0.870$\pm$0.011	&0.901$\pm$0.010	&0.929$\pm$0.042	&0.978$\pm$0.024	&0.940$\pm$0.035	\\
&MHIM-ABMIL~\cite{MHIM}	&0.872$\pm$0.015	&\Sd{0.902$\pm$0.023}	&0.920$\pm$0.010	&0.904$\pm$0.049	&0.979$\pm$0.022	&0.920$\pm$0.038	\\
&MHIM-Trans~\cite{MHIM}	&0.859$\pm$0.026	&0.888$\pm$0.019	&0.912$\pm$0.017	&0.880$\pm$0.150	&0.947$\pm$0.085	&0.890$\pm$0.142	\\
&ILRA~\cite{2023ILRA_MIL}  	&0.864$\pm$0.022	&0.896$\pm$0.017	&0.916$\pm$0.014	&0.892$\pm$0.105	&0.965$\pm$0.057	&0.900$\pm$0.103	\\
&SSMMIL~\cite{23MICCAIS4MIL_mamba}	&0.869$\pm$0.012	&0.900$\pm$0.010	&0.919$\pm$0.010	&0.917$\pm$0.072	&0.975$\pm$0.024	&0.930$\pm$0.059	\\
&ACMIL~\cite{zhang2023ACMIL}	&0.869$\pm$0.023	&0.890$\pm$0.038	&0.920$\pm$0.015	&\Sd{0.929$\pm$0.042}	&0.978$\pm$0.024	&\Sd{0.940$\pm$0.035}	\\
&WIKG~\cite{2024CVPRWIKG} 	&0.866$\pm$0.018	&0.894$\pm$0.018	&0.918$\pm$0.011	&0.892$\pm$0.093	&0.949$\pm$0.063	&0.900$\pm$0.093	\\
&MambaMIL~\cite{Hao2024mambamil}	&0.862$\pm$0.021	&0.887$\pm$0.022	&0.916$\pm$0.013	&0.891$\pm$0.053	&0.977$\pm$0.018	&0.910$\pm$0.041	\\
&PAMoE~\cite{wu2025learningcvpr} &0.869$\pm$0.015	&0.899$\pm$0.006	&0.919$\pm$0.009  &0.892$\pm$0.087	&0.955$\pm$0.071	&0.911$\pm$0.066\\
\rowcolor{tablecolor} 
&\textbf{AINet (Ours)}	&\Ft{0.880$\pm$0.021}	&\Ft{0.909$\pm$0.018}	&\Ft{0.926$\pm$0.014}	&\Ft{0.937$\pm$0.080}	&\Ft{0.982$\pm$0.039}	&\Ft{0.945$\pm$0.071}	\\
\bottomrule
\end{tabular}
\label{Table1_brca_esca}
\vspace{-4mm}
\end{table*}

\section{Experiments}
\label{Experiments}

To comprehensively evaluate our AINet, we integrate it into representative MIL predictors and conduct extensive experiments on cancer prediction tasks. We also perform ablation studies to assess the contributions of key components, including the DAM and ARC modules. Furthermore, we examine the effects of the AINet hyperparameter settings.

\begin{table*}[!ht] \centering
\caption{Quantitative results of AINet compared with representative MIL baselines on the BRACS dataset.} 
\setlength{\tabcolsep}{3.8pt} 
\renewcommand{\arraystretch}{1}
\newcommand{\Ft}[1]{\textcolor{red}{\textbf{#1}}}
\newcommand{\Sd}[1]{\textcolor{blue}{\textbf{#1}}}
\label{Table2_bracs}
\resizebox{\textwidth}{!}{ 
\begin{tabular}{l*{3}{c}*{3}{c}} %*{2}{l}
\toprule
% \multicolumn{1}{c}
{\multirow{3}{*}{Methods}}& \multicolumn{3}{c}{ResNet18} & \multicolumn{3}{c}{PLIP}  \\
\cmidrule(lr){2-4}  \cmidrule(lr){5-7}  %\cmidrule(lr){8-9} 
\multicolumn{1}{c}{} & Accuracy & AUC &F1 & Accuracy & AUC &F1 \\ %& FLOPs& Parameters
\midrule
ABMIL~\cite{ABMIL}	&0.691$\pm$0.041	&0.816$\pm$0.020	&0.604$\pm$0.055	&0.754$\pm$0.042	&0.875$\pm$0.036	&0.705$\pm$0.044	\\
DSMIL~\cite{CVPR21DSMIL}	&0.657$\pm$0.026	&0.795$\pm$0.021	&0.555$\pm$0.016	&0.717$\pm$0.044	&0.847$\pm$0.039	&0.629$\pm$0.041	\\
CLAM~\cite{CLAM}	&0.689$\pm$0.036	&0.830$\pm$0.028	&0.601$\pm$0.024	&0.742$\pm$0.054	&0.871$\pm$0.023	&0.692$\pm$0.043	\\
TransMIL~\cite{Transmil}	&0.706$\pm$0.044	&0.799$\pm$0.035	&0.596$\pm$0.036	&0.709$\pm$0.032	&0.851$\pm$0.022	&0.640$\pm$0.059	\\
DTFD-MaxM~\cite{2022DTFDMIL_CVPR}	&0.698$\pm$0.030	&0.815$\pm$0.062	&0.610$\pm$0.044	&0.724$\pm$0.054	&0.838$\pm$0.063	&0.617$\pm$0.075	\\
DTFD-AFS~\cite{2022DTFDMIL_CVPR}	&0.676$\pm$0.056	&0.823$\pm$0.039	&0.614$\pm$0.054	&0.759$\pm$0.025	&0.872$\pm$0.034	&0.669$\pm$0.037	\\
MHIM-ABMIL~\cite{MHIM}	&0.715$\pm$0.035	&\Sd{0.851$\pm$0.033}	&0.624$\pm$0.039	&0.765$\pm$0.034	&\Sd{0.881$\pm$0.038}	&0.709$\pm$0.055	\\
MHIM-Trans~\cite{MHIM}	&0.689$\pm$0.026	&0.833$\pm$0.028	&0.613$\pm$0.016	&0.717$\pm$0.046	&0.859$\pm$0.028	&0.669$\pm$0.045	\\
ILRA~\cite{2023ILRA_MIL}	
&0.702$\pm$0.037	&0.851$\pm$0.043	&\Sd{0.642$\pm$0.052} &0.717$\pm$0.025	&0.852$\pm$0.029	&0.622$\pm$0.039	\\
SSMMIL~\cite{23MICCAIS4MIL_mamba}	&\Sd{0.721$\pm$0.037}	&0.834$\pm$0.028	&0.620$\pm$0.048	&0.723$\pm$0.027	&0.867$\pm$0.031	&0.679$\pm$0.011	\\
ACMIL~\cite{zhang2023ACMIL}	&0.698$\pm$0.041	&0.849$\pm$0.044	&0.633$\pm$0.044	&\Sd{0.767$\pm$0.040}	&0.872$\pm$0.035	&\Sd{0.710$\pm$0.029}\\
WIKG~\cite{2024CVPRWIKG}	&0.706$\pm$0.053	&0.837$\pm$0.030	&0.637$\pm$0.055	&0.724$\pm$0.032	&0.858$\pm$0.018	&0.662$\pm$0.049	\\
MambaMIL~\cite{Hao2024mambamil}	&0.706$\pm$0.066	&0.834$\pm$0.039	&0.636$\pm$0.071	&0.736$\pm$0.049	&0.875$\pm$0.038	&0.670$\pm$0.069	\\
PAMoE~\cite{wu2025learningcvpr} &0.661$\pm$0.057	&0.800$\pm$0.046	&0.593$\pm$0.035 &0.730$\pm$0.032	&0.850$\pm$0.023	&0.677$\pm$0.029\\
\rowcolor{tablecolor} 
\textbf{AINet (Ours)} 
& \Ft{0.745$\pm$0.060}	&\Ft{0.864$\pm$0.051}	& \Ft{0.687$\pm$0.076}
&\Ft{0.771$\pm$0.040}	&\Ft{0.890$\pm$0.032}	&\Ft{0.723$\pm$0.057}\\
\bottomrule
\end{tabular}
}
\vspace{-4mm}
\end{table*}

\subsection{Datasets and Implementation Details}
We evaluate the performance of AINet on three public benchmarks and employ two widely used feature extractors: ResNet-18~\cite{ResNet18} pretrained on ImageNet, and PLIP~\cite{huang2023visualPLIP}, a foundation model pretrained on large-scale multimodal pathology data. In brief, the used datasets are summarized as follows:
\textbf{(1) TCGA Breast Cancer Dataset (BRCA).} 
This dataset comprises 952 WSIs from the BRCA project, including 749 invasive ductal carcinoma (IDC) and 203 invasive lobular carcinoma (ILC) cases. Following prior works~\cite{liu2023pseudoMixup}, the dataset is divided into 65\% for training, 10\% for validation, and 25\% for testing. 
\textbf{(2) TCGA Esophageal Cancer Dataset (ESCA).} 
ESCA contains 156 diagnostic WSIs, including 90 squamous cell carcinoma and 66 adenocarcinoma cases~\cite{TCGALung}. Data are divided into training, validation, and test sets with a 3:1:1 ratio, following~\cite{MuRCL}.
\textbf{(3) Breast Carcinoma Subtyping Dataset (BRACS).} 
BRACS consists of 265 benign, 89 atypical, and 193 malignant breast tumors~\cite{brancati2022bracs}. Following the official data split~\cite{brancati2022bracs,zhang2023ACMIL}, 395 WSIs are used for training, 65 for validation, and 87 for testing.

All WSIs are preprocessed following the CLAM~\cite{CLAM} pipeline to extract non-background tissue 256$\times$256 patches at 10× magnification. The model is optimized using bag-level cross-entropy loss and the AdamW optimizer~\cite{Adamm}, with a weight decay of $1e-5$ and an initial learning rate of $1e-4$. We train the AINet model for 100 epochs with a batch size of 1 (\emph{i.e.,} one bag per batch) on a single NVIDIA RTX 4090 GPU.

\subsection{Baseline and Evaluation Metrics}
\label{sub:Dataset_and_Metrics}
\noindent{\textbf{Baseline.}} We compare our AINet with 14 representative MIL methods, including attention-based approaches (ABMIL~\cite{ABMIL}, CLAM~\cite{CLAM}, DSMIL~\cite{CVPR21DSMIL}, MHIM-ABMIL~\cite{MHIM},  ACMIL~\cite{zhang2023ACMIL}, DTFD-MaxM~\cite{2022DTFDMIL_CVPR}, DTFD-AFS~\cite{2022DTFDMIL_CVPR}), Transformer-based methods (TransMIL~\cite{Transmil},  MHIM-TransMIL~\cite{MHIM}, ILRA~\cite{2023ILRA_MIL}, PAMoE~\cite{wu2025learningcvpr}), graph-based approaches WIKG~\cite{2024CVPRWIKG}, and Mamba-based methods (SSMMIL~\cite{23MICCAIS4MIL_mamba}, MambaMIL~\cite{Hao2024mambamil}). All baselines are implemented using their official repositories.

\noindent{\textbf{Evaluation Metrics.}}  
Following prior works~\cite{CVPR21DSMIL, IBMIL, cvpr24PAMIL}, we evaluate prediction performance using standard metrics, including the area under the receiver operating characteristic curve (AUC), accuracy, and F1 score, with a fixed threshold of 0.5. To eliminate evaluation bias, we follow the 5-fold cross-validation setting adopted in previous works~\cite{MHIM, Hao2024mambamil,TMI23bayesian_pseudolabel_thre}. In each fold, training and validation sets are divided according to predefined dataset ratios for hyperparameter tuning. The final results are reported as the mean and standard deviation across all five test folds.

\begin{table*}[!htb]
\centering
\setlength{\tabcolsep}{4pt} 
\renewcommand{\arraystretch}{1.1}
\caption{Validation on the TCGA-ESCA dataset. The \textcolor{imp}{red} denotes performance improvement, while \textcolor{impgre}{green} indicates reduced complexity.} 
\resizebox{\textwidth}{!}{
\begin{tabular}{@{}lllllllll@{}}
\toprule
\multirow{2}{*}{Methods} & \multicolumn{3}{c}{ResNet18} & \multicolumn{3}{c}{PLIP} & \multirow{2}{*}{FLOPs (G)} & \multirow{2}{*}{\#Param. (M)} \\
\cmidrule(lr){2-4} \cmidrule(lr){5-7} 
 &  Accuracy & AUC & F1 & Accuracy & AUC & F1 & & \\
\midrule
TransMIL~\cite{Transmil}	&0.815$\pm$0.081	&0.878$\pm$0.086	&0.847$\pm$0.065	&0.904$\pm$0.062	&0.988$\pm$0.018	&0.920$\pm$0.051	
&84.22	&0.68	\\
+AINet	&0.822$\pm$0.121\textcolor{imp}{(0.7)}	
&0.889$\pm$0.116\textcolor{imp}{(1.1)}	&0.8438$\pm$0.108\textcolor{imp}{(0.4)}	&0.916$\pm$0.054\textcolor{imp}{(1.2)}	&0.993$\pm$0.006\textcolor{imp}{(0.5)}	&0.923$\pm$0.052\textcolor{imp}{(0.3)}	&22.30\textcolor{impgre}{(61.92)}	&0.19\textcolor{impgre}{(0.49)}	\\
ILRA~\cite{2023ILRA_MIL}	&0.803$\pm$0.136	&0.896$\pm$0.097	&0.838$\pm$0.116	&0.892$\pm$0.105	&0.965$\pm$0.057	&0.900$\pm$0.103	&123.05	&1.85	\\
+AINet	
&0.866$\pm$0.098\textcolor{imp}{(6.3)}	&0.915$\pm$0.091\textcolor{imp}{(1.9)}	&0.882$\pm$0.090\textcolor{imp}{(4.4)}	&0.905$\pm$0.099\textcolor{imp}{(1.3)}	&0.967$\pm$0.054\textcolor{imp}{(0.2)}	&0.918$\pm$0.083\textcolor{imp}{(1.8)}	&21.86\textcolor{impgre}{(101.19)}	&0.19\textcolor{impgre}{(1.66)}	\\
WIKG~\cite{2024CVPRWIKG}	&0.770$\pm$0.131	&0.874$\pm$0.106	&0.793$\pm$0.121	&0.892$\pm$0.093	&0.949$\pm$0.063	&0.900$\pm$0.093	&49.80	&0.43	\\
+AINet	&0.827$\pm$0.066\textcolor{imp}{(5.7)}	
&0.906$\pm$0.070\textcolor{imp}{(3.2)}	&0.850$\pm$0.057\textcolor{imp}{(5.7)}	&0.943$\pm$0.068\textcolor{imp}{(5.1)}	&0.974$\pm$0.036\textcolor{imp}{(2.5)}	&0.949$\pm$0.061\textcolor{imp}{(4.9)}	&21.88\textcolor{impgre}{(27.92)}	&0.18\textcolor{impgre}{(0.25)}	\\
SSMMIL~\cite{23MICCAIS4MIL_mamba}	&0.783$\pm$0.111	&0.892$\pm$0.081	&0.823$\pm$0.096	&0.917$\pm$0.072	&0.985$\pm$0.024	&0.930$\pm$0.059	&92.14	&0.79	\\
+AINet	&0.828$\pm$0.122\textcolor{imp}{(4.5)}
&0.903$\pm$0.093\textcolor{imp}{(1.1)}	&0.865$\pm$0.091\textcolor{imp}{(4.2)}	&0.930$\pm$0.067\textcolor{imp}{(1.3)}	&0.985$\pm$0.033\textcolor{imp}{(0.0)}
&0.938$\pm$0.060\textcolor{imp}{(0.8)}	&21.74\textcolor{impgre}{(70.40)}	&0.17\textcolor{impgre}{(0.62)}	\\
\bottomrule
\label{MIL+AINet}
\end{tabular}
}
\vspace{-5mm}
\end{table*}

\subsection{Comparison with State-of-the-Art Methods}
\label{sub:ComparisonSOTA}
The comparison results on two benchmark WSI datasets 
are summarized in Table~\ref{Table1_brca_esca}. We can find that our AINet consistently outperforms all baselines across all evaluation metrics. Specifically, averaged across the TCGA-BRCA and TCGA-ESCA datasets, AINet surpasses bag-level methods such as TransMIL~\cite{Transmil}, MambaMIL~\cite{Hao2024mambamil}, and PAMoE~\cite{wu2025learningcvpr} by 3.6\%, 2.2\%, and 3.9\% in accuracy, and by 3.7\% and 1.3\%, and 3.3\%in AUC, respectively. Compared with attention-based instance selection methods, including CLAM~\cite{CLAM} and ACMIL~\cite{zhang2023ACMIL}, AINet achieves average gains of 2.3\% and 0.88\% in accuracy, and 1.6\% and 0.95\% in AUC, respectively. In addition, AINet outperforms the pseudo-bag-based DTFD~\cite{2022DTFDMIL_CVPR}, achieving average improvements of 3.4\%, 4.1\%, and 2.3\% in accuracy, AUC, and F1, respectively.
These superior performances largely stem from the proposed anchor instance learning, which provides high-quality and discriminative inputs for the MIL predictor. In contrast, instance selection and fusion methods are inevitably affected by regional heterogeneity, thereby limiting their representational quality.

To further validate AINet on multi-class classification tasks, we evaluate it on the BRACS dataset. As listed in Table~\ref{Table2_bracs}, AINet achieves consistent performance gains, surpassing ILRA~\cite{2023ILRA_MIL}, SSMMIL~\cite{23MICCAIS4MIL_mamba}, and WIKG~\cite{2024CVPRWIKG} by 4.9\%, 3.6\%, and 4.3\% in average accuracy, respectively. Furthermore, although MHIM~\cite{MHIM} leverages knowledge distillation to identify informative instances and achieves a relatively higher AUC, its accuracy and F1 drop by 3.6\% and 5.1\%, respectively, due to the sparse discriminative information in BRACS. In contrast, AINet effectively alleviates this issue through dual-level anchor mining and anchor-guided region correction, enabling concise yet accurate representation of such complex instance distributions, and achieving improvements of 1.8\%, 1.1\%, and 3.9\% in accuracy, AUC, and F1, respectively. Notably, our AINet achieves significant reductions in FLOPs and the number of parameters, as shown in Figure~\ref{fig_FLOPs}.

\subsection{AINet Generalizablity}
To further demonstrate the generalization and robustness of the proposed AINet, we integrate it into four representative MIL predictors on the TCGA-ESCA dataset. As reported in Table~\ref{MIL+AINet}, AINet could improve the performance of all MIL models while saving computational resources. When combined with AINet, the graph-based WIKG~\cite{2024CVPRWIKG} and the SSM-based SSMMIL~\cite{23MICCAIS4MIL_mamba} achieve average accuracy gains of 5.4\% and 2.9\%, and F1 improvements of 5.3\% and 2.5\%, respectively, while reducing FLOPs by 57\% and 76.4\%. Moreover, through cross-region interaction while eliminating redundancy, AINet reduces FLOPs by 73.5\% and 82.08\%, and parameters by 72.05\% and 89.75\%, when compared to the multi-head attention–based TransMIL~\cite{Transmil} and ILRA~\cite{2023ILRA_MIL}, respectively.

\begin{table*}[!t]
\centering
\vspace{-3mm}
\setlength{\tabcolsep}{2pt} 
\renewcommand{\arraystretch}{1} 
\caption{Ablation study of the key components on the TCGA-ESCA and BRACS datasets with ResNet18 features. The \colorbox{tablecolor}{gray} highlights our results. We set $L=4$ regions, select the top $k=20\%$ instances as AIs, and apply a masking ratio of $r=90\%$ within each region.}
\resizebox{\textwidth}{!}{ 
\begin{tabular}{lccccccccc}
\toprule
{\multirow{3}{*}{Model}} &{\multirow{3}{*}{DAM}} &{\multirow{3}{*}{ACF}} &{\multirow{3}{*}{ARC}} & \multicolumn{3}{c}{TCGA-ESCA}  & \multicolumn{3}{c}{BRACS} \\
\cmidrule(lr){5-7} \cmidrule(lr){8-10} 
& & & & Accuracy & AUC & F1 & Accuracy & AUC & F1 \\
\midrule
Baseline & \ding{55} & \ding{55} & \ding{55} &0.777$\pm$0.134 &0.866$\pm$0.095	&0.808$\pm$0.114	&0.691$\pm$0.056	&0.809$\pm$0.054	&0.602$\pm$0.073\\

\textit{w} DAM & \ding{51} & \ding{55} & \ding{55} &0.809$\pm$0.117	&0.886$\pm$0.128	&0.841$\pm$0.095	&0.713$\pm$0.047	&0.835$\pm$0.040	&0.594$\pm$0.061\\

\textit{w} DAM+MHA & \ding{51} & \ding{55} & \ding{55}  &0.839$\pm$0.088	&0.935$\pm$0.064	&0.867$\pm$0.069	&0.719$\pm$0.061	&0.845$\pm$0.041	&0.624$\pm$0.063\\

\textit{w} DAM+ACF & \ding{51} & \ding{51} & \ding{55} &0.840$\pm$0.108	&0.938$\pm$0.073	&0.868$\pm$0.083	&0.732$\pm$0.039	&0.844$\pm$0.036	&0.631$\pm$0.045\\
\rowcolor{tablecolor} 
\textit{w} DAM+ARC (AINet) &\ding{51} & \ding{51} & \ding{51} &0.873$\pm$0.098	&0.946$\pm$0.046	&0.894$\pm$0.080	&0.737$\pm$0.025	&0.854$\pm$0.046	&0.638$\pm$0.040\\

\bottomrule
\end{tabular}
}
\label{Ablationstudy}
\vspace{-4mm}
\end{table*}

\begin{table*}[!t]
\centering
\setlength{\tabcolsep}{5pt} 
\renewcommand{\arraystretch}{1}
\caption{Ablation study of the proposed DAM on the TCGA-ESCA and BRACS datasets using ResNet18 features. We set $L=4$ regions, select the top $k=20\%$ instances as AIs, and apply a masking ratio of $r=90\%$ within each region.}
\resizebox{\textwidth}{!}
{
\begin{tabular}{@{}lcccccccc@{}}
\toprule
\multirow{2}{*}{Model} &{\multirow{2}{*}{Bag-level}} &{\multirow{2}{*}{Region-level}} & \multicolumn{3}{c}{TCGA-ESCA} & \multicolumn{3}{c}{BRACS}\\
\cmidrule(lr){4-6} \cmidrule(lr){7-9} 
 && &  Accuracy & AUC & F1 & Accuracy & AUC & F1  \\
\midrule
Baseline & \ding{55} & \ding{55} &0.777$\pm$0.134 &0.866$\pm$0.095	&0.808$\pm$0.114	&0.691$\pm$0.056	&0.809$\pm$0.054	&0.602$\pm$0.073\\
\textit{w} Attention & \ding{55} & \ding{55} &0.816$\pm$0.156	&0.890$\pm$0.109	&0.839$\pm$0.131	&0.713$\pm$0.043	&0.837$\pm$0.041	&0.638$\pm$0.054	\\
\textit{w} Max-pooling & \ding{55} & \ding{55} &0.808$\pm$0.093	&0.918$\pm$0.093	&0.842$\pm$0.067	&0.681$\pm$0.040	&0.820$\pm$0.058	&0.593$\pm$0.057	\\
\textit{w} Bag-similar & \ding{51} & \ding{55} &0.828$\pm$0.133	&0.923$\pm$0.075	&0.862$\pm$0.099	&0.687$\pm$0.048	&0.816$\pm$0.040	&0.575$\pm$0.051	\\
\textit{w} Region-similar & \ding{55} & \ding{51} &0.834$\pm$0.140	&0.920$\pm$0.069	&0.869$\pm$0.107	&0.728$\pm$0.044	&0.835$\pm$0.047	&0.619$\pm$0.059	\\
% \rowcolor{tablecolor}
\rowcolor{tablecolor} 
\textit{w} DAM & \ding{51} & \ding{51} &0.873$\pm$0.098	 &0.946$\pm$0.046	&0.894$\pm$0.080	&0.737$\pm$0.025	&0.854$\pm$0.046	&0.638$\pm$0.040\\
\bottomrule
\end{tabular}
}
\label{AblationstudyDAM}
\vspace{-4mm}
\end{table*}

\subsection{Ablation Studies}%
\label{sub:Ablation}
\noindent{\textbf{Validation on Basic Components.}}
We conduct ablation experiments to evaluate the contribution of the key components in AINet, including DAM and ARC, to the overall performance. As a baseline, both DAM and ARC modules are removed, and regional instances are directly fed into an attention-based predictor for bag-level classification.
To assess the effectiveness of DAM, we compare it against several alternative instance selection strategies, as summarized in Table~\ref{AblationstudyDAM}.
Moreover, to examine the role of ARC in alleviating regional heterogeneity and enhancing cross-region interaction, we design four variants: \textit{w} DAM, \textit{w} DAM+MHA, \textit{w} DAM+ACF, and \textit{w} DAM+ARC (AINet), where \textit{w} DAM+others denotes integrating the selected AIs into regions using different fusion mechanisms. Among them, \textit{w} DAM+ACF does not use the neighbor region for interaction during correction.

The quantitative results in Table~\ref{Ablationstudy} demonstrate that our complete AINet substantially outperforms its incomplete variants. Specifically, the \textit{w} DAM effectively identifies representative and discriminative instances through a local-global selection mechanism, yielding average gains of 2.7\% in accuracy and 2.3\% in AUC over the Baseline, respectively.
Moreover, the AI-guided fusion (\textit{w} DAM+ACF) further enhances regional representation quality, achieving 2.5\%, 3.05\%, and 3.2\% improvements in accuracy, AUC, and F1, respectively, compared to \textit{w} DAM. 
Although using informative AIs increases accuracy by 1.3\% on the BRACS dataset, the vanilla multi-head attention fusion (\textit{w} DAM+MHA) still underperforms our \textit{w} DAM+ACF due to the lack of explicit AIs guidance. Notably, the proposed \textit{w} DAM+ARC effectively mitigates regional heterogeneity and enhances cross-region interactions, resulting in significant improvements of 7.1\%, 6.25\%, and 6.1\% in accuracy, AUC, and F1 over the Baseline.

\noindent\textbf{DAM Variants.}
To further assess the effectiveness of the proposed DAM, Table~\ref{AblationstudyDAM} reports quantitative comparisons with several alternative instance selection strategies, including attention-based weighting (\textit{w} Attention), max-pooling (\textit{w} Max-Pooling), bag-level similarity weighting (\textit{w} Bag-similar, using $\cos(f_l^z, f_\text{bag})$ in Eq.~\ref{eq:similarity_weight}), and region-level similarity weighting (\textit{w} Region-similar, using $\cos(f_l^z, f_\text{reg}^l)$ in Eq.~\ref{eq:similarity_weight}).
Our DAM consistently outperforms all competing variants. Notably, selecting informative instances only at the bag-level leads to decreased accuracy: \textit{w} Bag-similar (-2.6\%), \textit{w} Max-pooling (-3.65\%), and \textit{w} Attention (-1.65\%) compared with \textit{w} Region-similar (0.781). In contrast, \textit{w} DAM achieves gains of 2.4\%, 2.25\%, and 2.2\% in Accuracy, AUC, and F1, respectively, indicating that our DAM effectively identifies representative and discriminative instances, leading to better bag predictions.

\begin{figure}[!t] \centering
\includegraphics[width=0.48\textwidth]{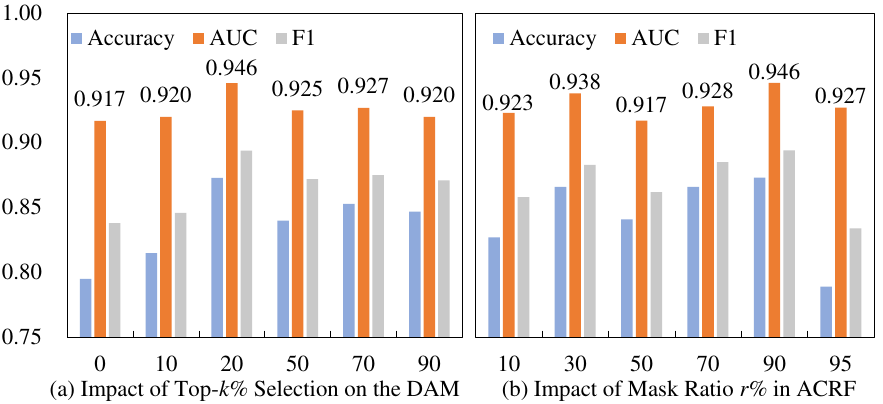}
\caption{Hyperparameter effects on the TCGA-ESCA dataset.} \label{fig_Hyperarameter}
\vspace{-5mm}
\end{figure}

\noindent{\textbf{Hyperparameter Analysis.}}
To analyze the influence of the number of AIs ($T = k\% \times N$) and the mask ratio ($r$), we conduct extensive experiments on AINet. As illustrated in Figure~\ref{fig_Hyperarameter}, appropriately setting $k\%$ and $r\%$ significantly enhances classification performance. This improvement primarily arises from retaining informative AIs, which alleviates tumor sparsity and regional heterogeneity, thereby yielding gains of 2.9\%, 7.8\%, and 5.6\% gain in accuracy, AUC, and F1 over the $k=0\%$ setting. Moreover, incorporating the masking scheme in ARC further enhances performance, with 2.3\% and 3.6\% improvements in accuracy and F1, respectively, compared with the $r=10\%$ setting.

\section{Conclusion}%
\label{Conclusion}
In this work, we proposed AINet, a novel MIL framework for WSI analysis. AINet introduces the concept of AIs, a compact set of informative instances that are locally representative and globally discriminative, to mitigate the regional heterogeneity. The DAM module identifies AIs by jointly considering instance similarities with both local region and global bag embeddings, while the ARC module leverages AIs to explicitly guide interactions across heterogeneous regions, correcting the non-discriminative patterns. Extensive experiments on widely used benchmark datasets demonstrate that AINet consistently outperforms SOTA MIL methods with fewer FLOPs and parameters, and can be seamlessly integrated into other predictors.

\clearpage
{\small
\bibliographystyle{ieee_fullname}
\bibliography{refs}

@String(CVPR= {IEEE Conf. Comput. Vis. Pattern Recog.})

@String(ICCV= {Int. Conf. Comput. Vis.})

@String(ECCV= {Eur. Conf. Comput. Vis.})

@String(ICLR = {Int. Conf. Learn. Represent.})

@String(PR   = {Pattern Recognition})

@String(AAAI = {AAAI})

@String(CVPR  = {CVPR})

@String(ICCV  = {ICCV})

@String(ECCV  = {ECCV})

@String(ICLR  = {ICLR})

@String(PR = {PR})

@inproceedings{tan2022fedproto,
  title={Fedproto: Federated prototype learning across heterogeneous clients},
  author={Tan, Yue and Long, Guodong and Liu, Lu and Zhou, Tianyi and Lu, Qinghua and Jiang, Jing and Zhang, Chengqi},
  booktitle={AAAI},
  volume={36},
  number={8},
  pages={8432--8440},
  year={2022}
}

@article{rothenhausler2021anchor,
  title={Anchor regression: Heterogeneous data meet causality},
  author={Rothenh{\"a}usler, Dominik and Meinshausen, Nicolai and B{\"u}hlmann, Peter and Peters, Jonas},
  journal={Journal of the Royal Statistical Society Series B: Statistical Methodology},
  volume={83},
  number={2},
  pages={215--246},
  year={2021},
  publisher={Oxford University Press}
}

@article{qin2024discriminative,
  title={Discriminative anchor learning for efficient multi-view clustering},
  author={Qin, Yalan and Pu, Nan and Wu, Hanzhou and Sebe, Nicu},
  journal={IEEE Transactions on Multimedia},
  year={2024},
  publisher={IEEE}
}

@inproceedings{ResNet18,
  author       = {Kaiming He and
                  Xiangyu Zhang and
                  Shaoqing Ren and
                  Jian Sun},
  title        = {Deep Residual Learning for Image Recognition},
  booktitle    = {CVPR},
  pages        = {770--778},
  year         = {2016}
}

@inproceedings{2022DTFDMIL_CVPR,
  author       = {Hongrun Zhang and
                  Yanda Meng and
                  Yitian Zhao and
                  Yihong Qiao and
                  Xiaoyun Yang and
                  Sarah E. Coupland and
                  Yalin Zheng},
  title        = {{DTFD-MIL:} Double-Tier Feature Distillation Multiple Instance Learning
                  for Histopathology Whole Slide Image Classification},
  booktitle    = {CVPR},
  pages        = {18780--18790},
  year         = {2022}
}

@article{2019RNNMIL_Nat,
  title = {Clinical-Grade Computational Pathology Using Weakly Supervised Deep Learning on Whole Slide Images},
  author = {Campanella, Gabriele and Hanna, Matthew G. and Geneslaw, Luke and Miraflor, Allen and Werneck Krauss Silva, Vitor and Busam, Klaus J. and Brogi, Edi and Reuter, Victor E. and Klimstra, David S. and Fuchs, Thomas J.},
  year = {2019},
  journal = {Nat. Med},
  volume = {25},
  number = {8},
  pages = {1301--1309},
  publisher = {Nature Publishing Group US New York},
  isbn = {1078-8956}
}

@inproceedings{2020MSDAMIL_cvpr,
  title = {Multi-Scale Domain-Adversarial Multiple-Instance CNN for Cancer Subtype Classification with Unannotated Histopathological Images},
  booktitle = {CVPR},
  author = {Hashimoto, Noriaki and Fukushima, Daisuke and Koga, Ryoichi and Takagi, Yusuke and Ko, Kaho and Kohno, Kei and Nakaguro, Masato and Nakamura, Shigeo and Hontani, Hidekata and Takeuchi, Ichiro},
  year = {2020},
  pages = {3852--3861}
}

@inproceedings{ABMIL,
  title = {Attention-Based Deep Multiple Instance Learning},
  booktitle = {ICML},
  author = {Ilse, Maximilian and Tomczak, Jakub and Welling, Max},
  year = {2018},
  pages = {2127--2136},
  isbn = {2640-3498}
}

@article{CLAM,
  title = {Data-Efficient and Weakly Supervised Computational Pathology on Whole-Slide Images},
  author = {Lu, Ming Y. and Williamson, Drew FK and Chen, Tiffany Y. and Chen, Richard J. and Barbieri, Matteo and Mahmood, Faisal},
  year = {2021},
  journal = {Nat. Biomed. Eng},
  volume = {5},
  number = {6},
  pages = {555--570},
  publisher = {Nature Publishing Group UK London},
  isbn = {2157-846X}
}

@article{Transmil,
  title = {Transmil: Transformer Based Correlated Multiple Instance Learning for Whole Slide Image Classification},
  author = {Shao, Zhuchen and Bian, Hao and Chen, Yang and Wang, Yifeng and Zhang, Jian and Ji, Xiangyang},
  year = {2021},
  journal = {NeurIPS},
  volume = {34},
  pages = {2136--2147}
}

@inproceedings{MHIM,
  title={Multiple Instance Learning Framework with Masked Hard Instance Mining for Whole Slide Image Classification},
  author={Tang, Wenhao and Huang, Sheng and Zhang, Xiaoxian and Zhou, Fengtao and Zhang, Yi and Liu, Bo},
  booktitle={ICCV},
  pages={4078--4087},
  year={2023}
}

@inproceedings{CVPR21DSMIL,
  author       = "Bin, Li.
                  Yin Li,
                  Kevin W. Eliceiri",
  title        = {Dual-Stream Multiple Instance Learning Network for Whole Slide Image
                  Classification With Self-Supervised Contrastive Learning},
  booktitle    = {CVPR},
  pages        = {14318--14328},
  year         = {2021}
}

@article{MuRCL,
  title={Murcl: Multi-instance reinforcement contrastive learning for whole slide image classification},
  author={Zhu, Zhonghang and Yu, Lequan and Wu, Wei and Yu, Rongshan and Zhang, Defu and Wang, Liansheng},
  journal={TMI},
  year={2022},
  publisher={IEEE}
}

@inproceedings{RankMix,
  author       = {Yuan{-}Chih Chen and
                  Chun{-}Shien Lu},
  title        = {RankMix: Data Augmentation for Weakly Supervised Learning of Classifying
                  Whole Slide Images with Diverse Sizes and Imbalanced Categories},
  booktitle    = {CVPR},
  pages        = {23936--23945},
  year         = {2023}
}

@inproceedings{HGNN_graph,
  author       = {Tsai Hor Chan and
                  Fernando Julio Cendra and
                  Lan Ma and
                  Guosheng Yin and
                  Lequan Yu},
  title        = {Histopathology Whole Slide Image Analysis with Heterogeneous Graph Representation Learning},
  booktitle    = {CVPR},
  pages        = {15661--15670},
  year         = {2023}
}

@inproceedings{IBMIL,
  author       = {Tiancheng Lin and
                  Zhimiao Yu and
                  Hongyu Hu and
                  Yi Xu and
                  Chang Wen Chen},
  title        = {Interventional Bag Multi-Instance Learning On Whole-Slide Pathological Images},
  booktitle    = {CVPR},
  pages        = {19830--19839},
  year         = {2023}
}

@inproceedings{BackWSIfinetuing23,
  author       = {Honglin Li and
                  Chenglu Zhu and
                  Yunlong Zhang and
                  Yuxuan Sun and
                  Zhongyi Shui and
                  Wenwei Kuang and
                  Sunyi Zheng and
                  Lin Yang},
  title        = {Task-Specific Fine-Tuning via Variational Information Bottleneck for Weakly-Supervised Pathology Whole Slide Image Classification},
  booktitle    = {CVPR},
  pages        = {7454--7463},
  year         = {2023}
}

@article{ProtoDiv,
  title={ProtoDiv: Prototype-guided Division of Consistent Pseudo-bags for Whole-slide Image Classification},
  author={Yang, Rui and Liu, Pei and Ji, Luping},
  journal={arXiv:2304.06652},
  year={2023}
}

@inproceedings{Adamm,
  title={Decoupled Weight Decay Regularization},
  author={Loshchilov, Ilya and Hutter, Frank},
  booktitle={ICLR},
  year = {2018}
}

@article{TCGALung,
  title={Review The Cancer Genome Atlas (TCGA): an immeasurable source of knowledge},
  author={Tomczak, Katarzyna and Czerwi{\'n}ska, Patrycja and Wiznerowicz, Maciej},
  journal={Contemporary oncology onkologia},
  volume={2015},
  number={1},
  pages={68--77},
  year={2015},
  publisher={Termedia}
}

@article{shapleyval,
  title={Shapley values-enabled progressive pseudo bag augmentation for whole-slide image classification},
  author={Yan, Renao and Sun, Qiehe and Jin, Cheng and Liu, Yiqing and He, Yonghong and Guan, Tian and Chen, Hao},
  journal={TMI},
  year={2024},
  publisher={IEEE}
}

@article{liu2023pseudoMixup,
  title={Pseudo-Bag Mixup Augmentation for Multiple Instance Learning-Based Whole Slide Image Classification},
  author={Liu, Pei and Ji, Luping and Zhang, Xinyu and Ye, Feng},
  journal={TMI},
  year={2024},
  publisher={IEEE}
}

@article{brancati2022bracs,
  title={Bracs: A dataset for breast carcinoma subtyping in h\&e histology images},
  author={Brancati, Nadia and Anniciello, Anna Maria and Pati, Pushpak and Riccio, Daniel and Scognamiglio, Giosu{\`e} and Jaume, Guillaume and De Pietro, Giuseppe and Di Bonito, Maurizio and Foncubierta, Antonio and Botti, Gerardo and others},
  journal={Database},
  volume={2022},
  pages={baac093},
  year={2022},
  publisher={Oxford University Press UK}
}

@inproceedings{zhang2023ACMIL,
  title={Attention-challenging multiple instance learning for whole slide image classification},
  author={Zhang, Yunlong and Li, Honglin and Sun, Yunxuan and Zheng, Sunyi and Zhu, Chenglu and Yang, Lin},
  booktitle={ECCV},
  pages={125--143},
  year={2025},
}

@inproceedings{AAAI2024CIMIL_pseudolabel_thre,
  title={Boosting Multiple Instance Learning Models for Whole Slide Image Classification: A Model-Agnostic Framework Based on Counterfactual Inference},
  author={Lin, Weiping and Zhuang, Zhenfeng and Yu, Lequan and Wang, Liansheng},
  booktitle={AAAI},
  pages={3477--3485},
  year={2024}
}

@article{TMI23bayesian_pseudolabel_thre,
  title={Bayesian collaborative learning for whole-slide image classification},
  author={Yu, Jin-Gang and Wu, Zihao and Ming, Yu and Deng, Shule and Wu, Qihang and Xiong, Zhongtang and Yu, Tianyou and Xia, Gui-Song and Jiang, Qingping and Li, Yuanqing},
  journal={TMI},
  year={2023},
  publisher={IEEE}
}

@InProceedings{cvpr24PAMIL,
    author    = {Zheng, Tingting and Jiang, Kui and Yao, Hongxun},
    title     = {Dynamic Policy-Driven Adaptive Multi-Instance Learning for Whole Slide Image Classification},
    booktitle = {CVPR},
    month     = {June},
    year      = {2024},
    pages     = {8028-8037}
}

@inproceedings{Hao2024mambamil,
  title={Mambamil: Enhancing long sequence modeling with sequence reordering in computational pathology},
  author={Yang, Shu and Wang, Yihui and Chen, Hao},
  booktitle={MICCAI},
pages={296--306},
  year={2024}
}

@inproceedings{23MICCAIS4MIL_mamba,
  title={Structured state space models for multiple instance learning in digital pathology},
  author={Fillioux, Leo and Boyd, Joseph and Vakalopoulou, Maria and Courn{\`e}de, Paul-Henry and Christodoulidis, Stergios},
  booktitle={MICCAI},
  pages={594--604},
  year={2023}
}

@inproceedings{behrouz2024graphmamb_ACM,
  title={Graph mamba: Towards learning on graphs with state space models},
  author={Behrouz, Ali and Hashemi, Farnoosh},
  booktitle={ACM SIGKDD},
  pages={119--130},
  year={2024}
}

@article{quan2024globalPR,
  title={Global contrast-masked autoencoders are powerful pathological representation learners},
  author={Quan, Hao and Li, Xingyu and Chen, Weixing and Bai, Qun and Zou, Mingchen and Yang, Ruijie and Zheng, Tingting and Qi, Ruiqun and Gao, Xinghua and Cui, Xiaoyu},
  journal={PR},
  volume={156},
  pages={110745},
  year={2024},
  publisher={Elsevier}
}

@inproceedings{redundancy_cvpr2024MIL,
  title={Morphological prototyping for unsupervised slide representation learning in computational pathology},
  author={Song, Andrew H and Chen, Richard J and Ding, Tong and Williamson, Drew FK and Jaume, Guillaume and Mahmood, Faisal},
  booktitle={CVPR},
  pages={11566--11578},
  year={2024}
}

@inproceedings{2023ILRA_MIL,
  title={Exploring low-rank property in multiple instance learning for whole slide image classification},
  author={Xiang, Jinxi and Zhang, Jun},
  booktitle={ICLR},
  year={2023}
}

@inproceedings{2024CVPRWIKG,
  title={Dynamic graph representation with knowledge-aware attention for histopathology whole slide image analysis},
  author={Li, Jiawen and Chen, Yuxuan and Chu, Hongbo and Sun, Qiehe and Guan, Tian and Han, Anjia and He, Yonghong},
  booktitle={CVPR},
  pages={11323--11332},
  year={2024}
}

@article{huang2023visualPLIP,
  title={A visual--language foundation model for pathology image analysis using medical twitter},
  author={Huang, Zhi and Bianchi, Federico and Yuksekgonul, Mert and Montine, Thomas J and Zou, James},
  journal={Nat. Med.},
  volume={29},
  number={9},
  pages={2307--2316},
  year={2023},
  publisher={Nature Publishing Group US New York}
}

@article{zheng2025graphmamba,
  title={GraphMamba: Whole slide image classification meets graph-driven selective state space model},
  author={Zheng, Tingting and Yao, Hongxun and Zhao, Sicheng and Jiang, Kui and Xiao, Yi},
  journal={PR},
  pages={111768},
  year={2025},
  publisher={Elsevier}
}

@inproceedings{zheng2025oodml,
  title={OODML: Whole Slide Image Classification Meets Online Pseudo-Supervision and Dynamic Mutual Learning},
  author={Zheng, Tingting and Jiang, Kui and Yao, Hongxun and Xiao, Yi and Wang, Zhongyuan},
  booktitle={AAAI},
  volume={39},
  number={10},
  pages={10626--10634},
  year={2025}
}

@inproceedings{zheng2025gmmamba,
  title={GMMamba: Group Masking Mamba for Whole Slide Image Classification},
  author={Zheng, Tingting and Yao, Hongxun and Jiang, Kui and Xiao, Yi and Zhao, Sicheng},
  booktitle={ICCV},
  pages={9935--9944},
  year={2025}
}

@inproceedings{zheng2025m3amba,
  title={M3amba: Memory Mamba is All You Need for Whole Slide Image Classification},
  author={Zheng, Tingting and Jiang, Kui and Xiao, Yi and Zhao, Sicheng and Yao, Hongxun},
  booktitle={CVPR},
  pages={15601--15610},
  year={2025}
}

@article{zheng2023learning,
  title={Learning how to detect: A deep reinforcement learning method for whole-slide melanoma histopathology images},
  author={Zheng, Tingting and Chen, Weixing and Li, Shuqin and Quan, Hao and Zou, Mingchen and Zheng, Song and Zhao, Yue and Gao, Xinghua and Cui, Xiaoyu},
  journal={CMIG},
  volume={108},
  pages={102275},
  year={2023},
  publisher={Elsevier}
}

@inproceedings{shou2025graphiccv,
  title={Graph domain adaptation with dual-branch encoder and two-level alignment for whole slide image-based survival prediction},
  author={Shou, Yuntao and Cao, Xiangyong and Yan, Peiqiang and Hui, Qiao and Zhao, Qian and Meng, Deyu},
  booktitle={ICCV},
  pages={19925--19935},
  year={2025}
}

@inproceedings{ganguly2025mergegraphcvpr,
  title={MERGE: Multi-faceted Hierarchical Graph-based GNN for Gene Expression Prediction from Whole Slide Histopathology Images},
  author={Ganguly, Aniruddha and Chatterjee, Debolina and Huang, Wentao and Zhang, Jie and Yurovsky, Alisa and Johnson, Travis Steele and Chen, Chao},
  booktitle={CVPR},
  pages={15611--15620},
  year={2025}
}

@inproceedings{guo2025focuscvpr,
  title={Focus: Knowledge-enhanced adaptive visual compression for few-shot whole slide image classification},
  author={Guo, Zhengrui and Xiong, Conghao and Ma, Jiabo and Sun, Qichen and Feng, Lishuang and Wang, Jinzhuo and Chen, Hao},
  booktitle={CVPR},
  pages={15590--15600},
  year={2025}
}

@inproceedings{li2025advancingcvpr,
  title={Advancing Multiple Instance Learning with Continual Learning for Whole Slide Imaging},
  author={Li, Xianrui and Cui, Yufei and Li, Jun and Chan, Antoni B},
  booktitle={CVPR},
  pages={20800--20809},
  year={2025}
}

@inproceedings{wu2025learningcvpr,
  title={Learning Heterogeneous Tissues with Mixture of Experts for Gigapixel Whole Slide Images},
  author={Wu, Junxian and Chen, Minheng and Ke, Xinyi and Xun, Tianwang and Jiang, Xiaoming and Zhou, Hongyu and Shao, Lizhi and Kong, Youyong},
  booktitle={CVPR},
  pages={5144--5153},
  year={2025}
}

@inproceedings{zhang20252dmambacvpr,
  title={2dmamba: Efficient state space model for image representation with applications on giga-pixel whole slide image classification},
  author={Zhang, Jingwei and Nguyen, Anh Tien and Han, Xi and Trinh, Vincent Quoc-Huy and Qin, Hong and Samaras, Dimitris and Hosseini, Mahdi S},
  booktitle={CVPR},
  pages={3583--3592},
  year={2025}
}

@inproceedings{sun2025cpathcvpr,
  title={Cpath-omni: A unified multimodal foundation model for patch and whole slide image analysis in computational pathology},
  author={Sun, Yuxuan and Si, Yixuan and Zhu, Chenglu and Gong, Xuan and Zhang, Kai and Chen, Pingyi and Zhang, Ye and Shui, Zhongyi and Lin, Tao and Yang, Lin},
  booktitle={CVPR},
  pages={10360--10371},
  year={2025}
}

@inproceedings{raswa2025histofscvpr,
  title={HistoFS: Non-IID Histopathologic Whole Slide Image Classification via Federated Style Transfer with RoI-Preserving},
  author={Raswa, Farchan Hakim and Lu, Chun-Shien and Wang, Jia-Ching},
  booktitle={CVPR},
  pages={30251--30260},
  year={2025}
}

@inproceedings{li2024generalizablecvpr,
  title={Generalizable whole slide image classification with fine-grained visual-semantic interaction},
  author={Li, Hao and Chen, Ying and Chen, Yifei and Yu, Rongshan and Yang, Wenxian and Wang, Liansheng and Ding, Bowen and Han, Yuchen},
  booktitle={CVPR},
  pages={11398--11407},
  year={2024}
}

@inproceedings{shi2024vilacvpr,
  title={Vila-mil: Dual-scale vision-language multiple instance learning for whole slide image classification},
  author={Shi, Jiangbo and Li, Chen and Gong, Tieliang and Zheng, Yefeng and Fu, Huazhu},
  booktitle={CVPR},
  pages={11248--11258},
  year={2024}
}

@article{shao2023videoMIL,
  title={Video anomaly detection with NTCN-ML: A novel TCN for multi-instance learning},
  author={Shao, Wenhao and Xiao, Ruliang and Rajapaksha, Praboda and Wang, Mengzhu and Crespi, Noel and Luo, Zhigang and Minerva, Roberto},
  journal={PR},
  volume={143},
  pages={109765},
  year={2023},
}

@article{liu2018landmarkMIL,
  title={Landmark-based deep multi-instance learning for brain disease diagnosis},
  author={Liu, Mingxia and Zhang, Jun and Adeli, Ehsan and Shen, Dinggang},
  journal={MIA},
  volume={43},
  pages={157--168},
  year={2018},
}

@inproceedings{tang2024featureRRT,
  title={Feature re-embedding: Towards foundation model-level performance in computational pathology},
  author={Tang, Wenhao and Zhou, Fengtao and Huang, Sheng and Zhu, Xiang and Zhang, Yi and Liu, Bo},
  booktitle={CVPR},
  pages={11343--11352},
  year={2024}
}

@inproceedings{hou2016patch,
  title={Patch-based convolutional neural network for whole slide tissue image classification},
  author={Hou, Le and Samaras, Dimitris and Kurc, Tahsin M and Gao, Yi and Davis, James E and Saltz, Joel H},
  booktitle={CVPR},
  pages={2424--2433},
  year={2016}
}

@inproceedings{ge2024beyondanchor,
  title={Beyond prototypes: Semantic anchor regularization for better representation learning},
  author={Ge, Yanqi and Nie, Qiang and Huang, Ye and Liu, Yong and Wang, Chengjie and Zheng, Feng and Li, Wen and Duan, Lixin},
  booktitle={AAAI},
  volume={38},
  number={3},
  pages={1887--1895},
  year={2024}
}
}

\end{document}